\documentclass{article}
\usepackage{amsmath,amsfonts,amssymb,verbatim,float,anysize,setspace,fancyvrb,graphicx, multicol,mathrsfs, mathtools, colortbl, arydshln, multirow, varwidth}
\usepackage[utf8]{inputenc}
\usepackage[english]{babel}
\usepackage[numbers, super]{natbib} 
\usepackage{caption, mdframed}
\usepackage[top=.75in, bottom=.75in, left=.75in, right=.75in]{geometry}
\usepackage{authblk}

\parskip = \baselineskip
\title{Incorporating Contact Network Structure in \\ Cluster Randomized Trials}
\author[1]{Patrick C. Staples\thanks{email: \texttt{patrickstaples@fas.harvard.edu}}}
\author[2]{Elizabeth L. Ogburn\thanks{email: \texttt{eogburn@jhsph.edu}}}
\author[1]{Jukka-Pekka Onnela\thanks{email: \texttt{onnela@hsph.harvard.edu}}}
\affil[1]{\small{Department of Biostatistics, Harvard University, Boston, MA 02115, USA }}
\affil[2]{Department of Biostatistics, Johns Hopkins University, Baltimore MD, 21205, USA}

\date{}
\begin{document}
\maketitle
\def \resultswidth{6cm}
\renewcommand{\tabcolsep}{.1cm}
\definecolor{darkgreen}{rgb}{0.0, 0.5, 0.0}

\doublespacing
\textbf{Whenever possible, the efficacy of a new treatment, such as a drug or behavioral intervention, is investigated by randomly assigning some individuals to a treatment condition and others to a control condition, and comparing the outcomes between the two groups. Often, when the treatment aims to slow an infectious disease, groups or clusters of individuals are assigned en masse to each treatment arm.  The structure of interactions within and between clusters can reduce the power of the trial, i.e. the probability of correctly detecting a real treatment effect. We investigate the relationships among power, within-cluster structure, between-cluster mixing, and infectivity by simulating an infectious process on a collection of clusters. We demonstrate that current power calculations may be conservative for low levels of between-cluster mixing, but failing to account for moderate or high amounts can result in severely underpowered studies.  Power also depends on within-cluster network structure for certain kinds of infectious spreading. Infections that spread opportunistically through very highly connected individuals have unpredictable infectious breakouts, which makes it harder to distinguish between random variation and real treatment effects. Our approach can be used before conducting a trial to assess power using network information if it is available, and we demonstrate how empirical data can inform the extent of between-cluster mixing.}

\newpage
{\large\textbf{Introduction}}

In order to determine how effective a treatment is, it is common to randomly assign test subjects to different treatment arms.  In one arm, subjects receive the experimental treatment, and subjects in the other arm receive usual care or a placebo. Randomization helps to ensure that the treatment is the cause of any difference in outcomes between the subjects in the two treatment arms, as opposed to some pre-treatment characteristics of the individuals. If the treatment is effective, the probability that a trial will find a statistically significant difference attributed to the treatment is called the \emph{power} of the trial\cite{lipsey1990book}. Adequate power requires a sufficiently large number of subjects to be tested, which can be expensive or infeasible. Underpowered studies are not only less likely to find a true relationship if one exists, but they are also more likely to erroneously conclude that an effect exists when it does not\cite{button2013}\cite{ioannidis2005}. In order to control the probability of these errors, it is important to be able to accurately assess power before conducting a study.

When designing a randomized trial, we may not want or be able to randomly assign individuals to treatment. Individuals may be members of a cluster with complex  interactions, which makes it infeasible or unethical to assign some individuals within a cluster to treatment and others to control. For example, the spread of  HIV from infected to uninfected individuals in a small village might be slowed by offering its members information about safer sexual practices. In this case, it may be difficult or unethical to keep treated individuals' sex partners from sharing information or resources. We may instead choose to randomly select villages to participate in this regime, where villages correspond to naturally occurring clusters, and to compare HIV infection rates between treatment and control villages.  This type of experiment is called a \emph{Cluster Randomized Trial} (CRT)\cite{murray1998book} \cite{donner2000book} \cite{hayes2009book} \cite{eldridge2012book}.

The correlation in outcomes of individuals within a cluster (e.g. HIV infection statuses) is known to reduce the power of a trial\cite{donner2000book}. This correlation is generally summarized by a single parameter, called the \emph{Intracluster Correlation Coefficient}  (ICC)\cite{murray1998book}, which is the average pairwise correlation of outcomes within clusters.  This measure assumes that the correlation in outcomes for any two individuals within a cluster is identical. However, the structure of relationships within a cluster can be heterogeneous, and power may depend on that structure, which is not captured by the ICC. Usually, this structure is either ignored\cite{crespi2009} or analysis is performed using methods that allow it to be left unspecified\cite{wang2014}.  Furthermore, individuals are often likely to interact with others not only in the same cluster but also in other clusters, which can reduce the difference in outcomes between treated and untreated clusters, thereby decreasing power\cite{hemming2011}. For example, economic ties may exist between villages, the residents of which might then share information related to the treatment. If the treatment succeeds in slowing the infection rate in the treatment cluster, mixing between clusters will decrease the difference between outcomes in mixed clusters, so the power to detect a treatment effect will decrease and the probability of a false discovery will increase.  This must be addressed either by adding more clusters to the trial or increasing cluster sizes, both of which could be difficult and costly.  This issue is also often left unaddressed\cite{ukoumunne1999}\cite{hayes2000}\cite{carnegie2014}.

The effect of within-cluster structure and between-cluster mixing may depend on the type of infection spreading through each cluster. For example, a highly contagious infectious disease like the flu can spread more efficiently through more highly connected individuals\cite{zhou2006}. Other infectious diseases, such as a sexually transmitted disease, can only be transmitted to one person at a time, no matter how many partners one has. The number of individuals whom an infected person may infect at a given time is the person's \emph{infectivity}.  This quantity likely differs from person to person, and it depends crucially on the transmission dynamics of the disease.

\begin{figure}[H]
\centering
\includegraphics[width=16cm]{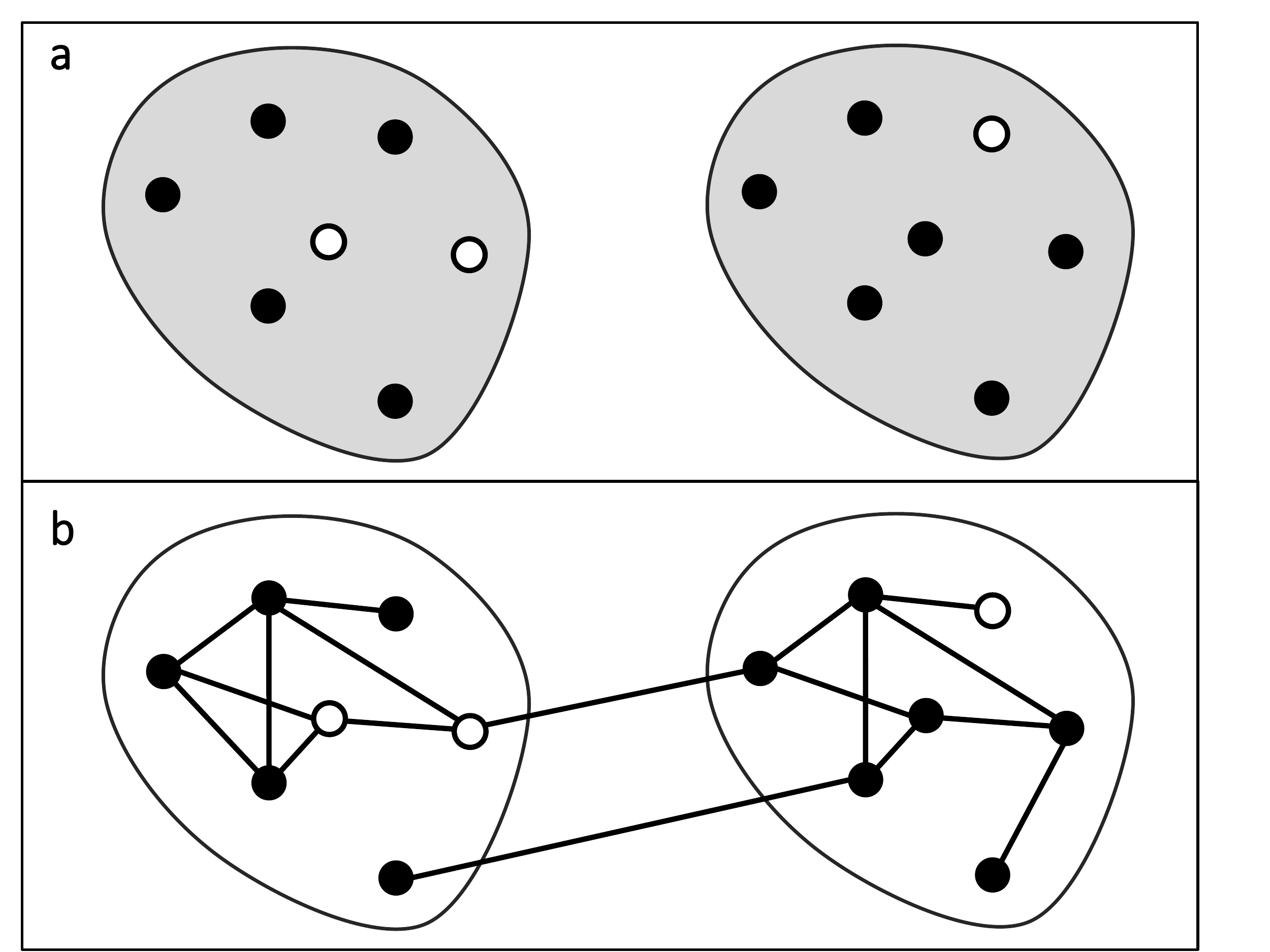}
\caption{A schematic comparing the Intracluster Correlation Coefficient (ICC) approach to the design of this study.  Each panel shows a cluster pair, and each enclosure represents a cluster.  Panel $\texttt{\textbf{a}}$ depicts cluster pair outcomes (circle colors) which are correlated (gray shading) within each cluster according to the ICC.  In contrast, Panel $\textbf{\texttt{b}}$ shows specific relationships (contact network ties) among individuals both within and between the two clusters, and outcomes among them will depend on an infection spreading only through these ties.  We show that modeling both contact network structure and the spreading process explicitly rather than modeling correlations across outcomes results in new findings about power in CRTs.}
\label{comparison}
\end{figure}

In this paper, we study, via simulation, the effect of within-cluster structure, the extent of between-cluster mixing, and infectivity on statistical power in CRTs.  We simulate the spread of an infectious process and investigate how power is affected by features of the process. Specifically, we consider two infections with different infectivities spreading through a collection of clusters.  We use a \emph{matched-pairs design}, wherein clusters in the study are paired, and each pair has one cluster assigned to treatment one to control\cite{eldridge2012book}.  We model the complex within-cluster correlation structure as a network in which edges represent possible transmission pathways between two individuals, comparing results across three different well-known network models.  We introduce a single parameter $\gamma$ that summarizes the extent of mixing between the two clusters comprising each cluster pair.  This approach departs from standard power calculations for CRTs, in which the researcher applies a formula that determines the required sample size as a function of the number and size of clusters, the ICC, and the effect size\cite{hayes1999}.  Figure \ref{comparison} depicts the different assumptions behind these two approaches. We show that our measure of mixing between clusters can have a strong effect on experimental \emph{power}, or the probability of correctly detecting a real treatment effect.  We also show that within-cluster structure can affect power for certain kinds of infectivity. We contrast this method to standard power calculations.  We end by demonstrating how to assess between-cluster mixing before designing a hypothetical CRT, using a network dataset of inter-regional cell phone calls.

{\large\textbf{Methods}}

We simulate both within-cluster structure and between-cluster mixing using network models.  We simulate pairs of clusters with each cluster in each pair initially generated as a stand-alone network.  We examine the Erd\"os-R\'enyi (ER)\cite{erdos1960}, Barab\'asi-Albert (BA)\cite{barabasi1999}, and stochastic blockmodel (SBM)\cite{anderson1992_1} random networks, and we simulate $2C$ clusters comprised of $n$ nodes each.  In order to explicitly allow for between-cluster mixing, we define a between-cluster mixing parameter $\gamma$ as the number of network edges between the treatment cluster and the control cluster, divided by the total number of edges in the cluster pair. To ensure that proportion $\gamma$ of the edges are shared across clusters, we perform degree-preserving rewiring\cite{newman2003_1} within each of the $C$ cluster-pairs until proportion $\gamma$ edges are shared between clusters.  We then use a compartmental model to simulate the spread of an infection across each cluster pair\cite{anderson1992book}.  All nodes are either susceptible ($S$) or infected ($I$), and nodes may only transition from $S$ to $I$.  The number of neighbors each node can potentially infect at any given time is called its \emph{infectivity}.  We consider both unit and degree infectivity, for which infected nodes may contact one or all of their neighbors at a given time, respectively.  Treated and control clusters infect their neighbors with equal probability under the null hypothesis, and infected individuals in treatment clusters infect with reduced probability under the alternative hypothesis.  Finally, we analyze the resulting trial under two different analysis scenarios, and we juxtapose our findings with a standard power calculation\cite{hayes1999}.  Table \ref{algorithm} summarizes our general simulation algorithm.  Next, we discuss each step in more detail.
\vspace{1cm}
\begin{table}[H]
\begin{center}
\begin{mdframed}[linewidth=1.25pt, leftmargin=1cm, rightmargin = 1.9cm, innertopmargin=.5cm]
\begin{tabular}{lll}
\large{\emph{1}} &\hspace{.5cm}\large{\textbf{Networks:}} &\hspace{1cm} \parbox[t]{8cm}{Generate $C$ cluster pairs using user-specified random networks.} \\[1cm]
\large{\emph{2}} &\hspace{.5cm}\large{\textbf{Mixing:}}    &\hspace{1cm} \parbox[t]{8cm}{Perform degree-preserving rewiring between the two clusters in each pair until proportion $\gamma$ ties are shared across them.} \\[1.5cm]
\large{\emph{3}} &\hspace{.5cm}\large{\textbf{Spreading:}} &\hspace{1cm} \parbox[t]{8cm}{Simulate a spreading process according to a suitable compartmental model.} \\[1cm]
\large{\emph{4}} &\hspace{.5cm}\large{\textbf{Analysis:}}  &\hspace{1cm} \parbox[t]{8cm}{Assess the empirical power of the simulation using the outcomes from the spreading process.} \\[.5cm]
\end{tabular}
\end{mdframed}
\caption{Our simulation algorithm used to assess the effect of within-cluster structure, between-cluster mixing and infectivity on statistical power.}
\label{algorithm}
\end{center}
\end{table}

\textbf{Networks.}\quad  Infectious disease dynamics have been studied extensively using deterministic ordinary differential equations \cite{keeling2007book} as well as network simulations\cite{pastorsatorras2014}. Using networks to simulate the spread of infection allows rich epidemic detail, and this added complexity facilitates exploration of the effect of cluster structure on power in CRTs. A brief treatment of these features using differential equations is in the supplement (\texttt{S1}).

A simple \emph{network} $\mathcal{G}$ consists of a set of $n$ \emph{nodes} (individuals) and a set of binary pairwise \emph{edges} (relationships) between the nodes. This structure can be compactly expressed by a symmetric  \emph{adjacency matrix} $\mathbf{A}_{n\times n}$. If an edge exists between individuals $i$ and $j$ then $A_{ij}=A_{ji}=1,$ and 0 otherwise. The \emph{degree} of node $i$, denoted by $k_{i},$ is the number of edges connecting node $i$ to other nodes in the network. Networks can be used to describe complex systems like social communities, the structure of metabolic pathways, and the World Wide Web; many reviews of this work are available\cite{newman2003} \cite{newman2010book} \cite{kolaczyk2009book} \cite{vanmiegham2014book}.

\begin{figure}[H]
\centering
\includegraphics[width=17.5cm]{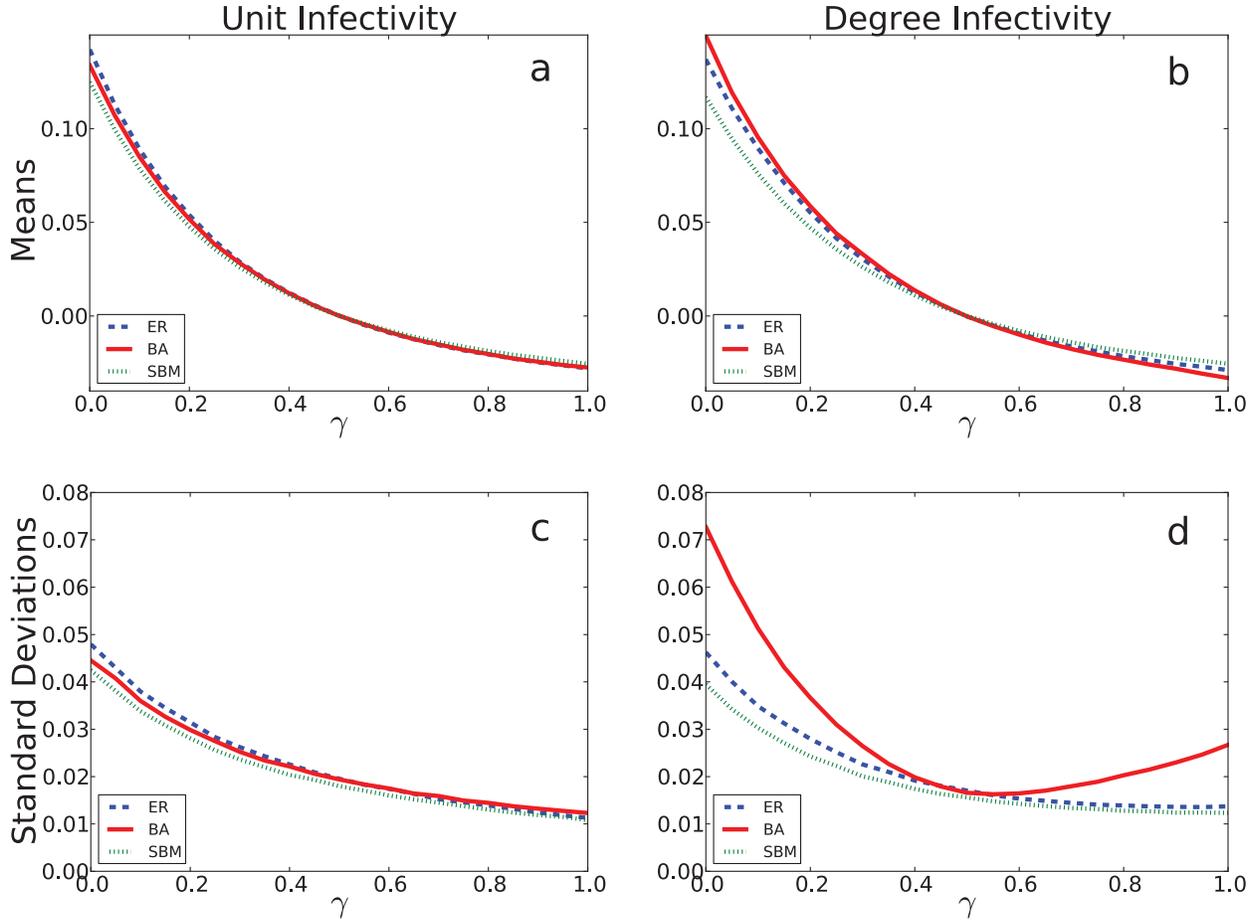}
\caption{The log risk ratio means and standard deviations under Scenario 1. The rows correspond to the means (Panels \texttt{\textbf{a}} and \texttt{\textbf{b}}) and standard deviations (Panels \texttt{\textbf{c}} and \texttt{\textbf{d}}), shown on the $y$ axis. The $x$-axis is the value of the mixing parameter $\gamma$, and each curve represents the three within-cluster network structures.  The left column shows the spread of an infection in which an infected node may only infect one neighbor per time step (unit infectivity), whereas the right column assumes one may spread an infection to each of their neighbors (degree infectivity).  We see that network topology has an effect on the variation of the log rate ratio only in the latter case.}
\label{metrics}
\end{figure}

\begin{figure}[H]
\centering
\includegraphics[width=17.5cm]{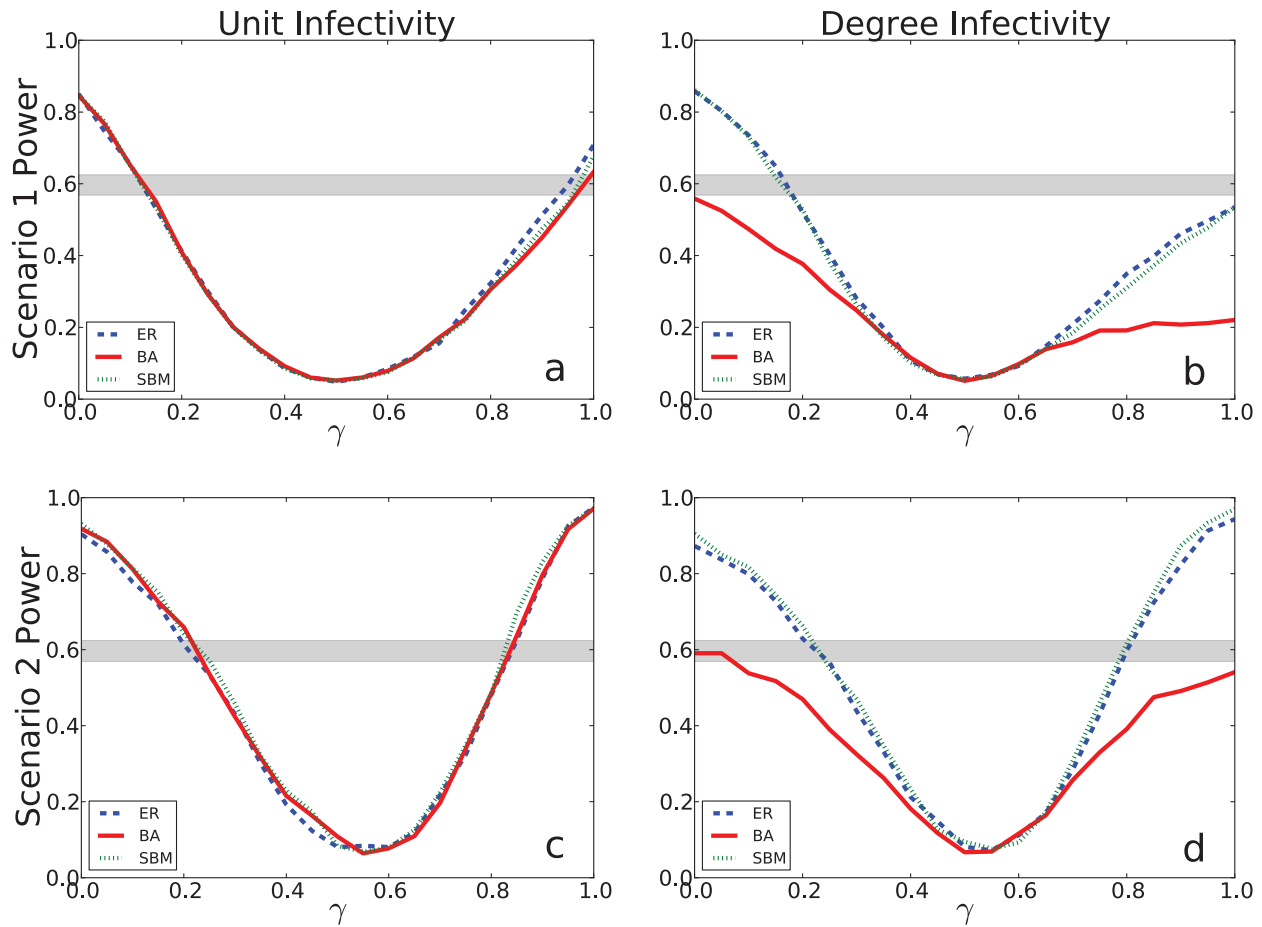}
\caption{Estimated power for each scenario. The blue, red, and green lines represent the ER, BA, and SBM network models, respectively.  The top row shows results for Scenario 1, and the bottom row shows results for Scenario 2.  The left column shows unit infectivity, and the right column shows degree infectivity.  The horizontal gray bars represent the expected power using the standard approach for a range of plausible values for the ICC.}
\label{power}
\end{figure}

A \emph{random network ensemble} is a collection of all possible networks specified either by a probability model or a mechanistic model\cite{newman2010book}. The simplest and  most studied random network is the Erd\"os-R\'enyi (ER) model\cite{erdos1960}, which assumes that each potential edge between any pair of nodes in a network occurs independently with unit probability. Nodes in an ER network tend to have degrees close to their shared expected value, while in real-world social and contact networks, the distribution of node degrees is typically \emph{heavy-tailed}: a few nodes are very highly connected (``hubs''), but most have small degree. To capture degree heterogeneity, we also simulate networks from the Barab\'asi-Albert (BA) model\cite{price1965}\cite{barabasi1999}. These networks are generated beginning with a small group of connected nodes and successively adding nodes one at a time, connecting them to the nodes in the existing network with probability proportional to the degree of each existing node. This mechanism has been shown to yield a power-law degree distribution:\cite{barabasi1999} $P(k)\sim k^{-\alpha}$ with $\alpha=3$. This distribution is heavy-tailed, so the probability that some individuals are highly connected is more likely than in other network models like the ER. While it can be difficult to assess whether an observed network has a power-law degree distribution\cite{clauset2009}, the BA model comes closer to capturing the heavy-tailed degree distributions observed in social networks than the ER model. Another hallmark of real-world social networks is that individuals tend to cluster together into \emph{communities}, or groups of individuals who share more edges with each other than between them\cite{porter2009}. We use \emph{stochastic blockmodels} (SBMs)\cite{anderson1992_1} to model within-cluster communities by assuming that each node is a member of a one block in a partition of blocks $\mathcal{B}$ comprising all nodes in the network, and that the probability of an edge between two nodes depends only on block membership (see supplementary material \texttt{S3} for additional details). Other popular families of random networks include Exponential Random Graphs (ERGMs)\cite{frank1986} and Small-World network of Watts and Strogatz, among others\cite{watts1998}. We leave their implications for CRTs for future research.  Network instances generated using Python's \texttt{networkx} library. Each node within each cluster has the same expected number of edges $\langle k \rangle=4$.  For Figures \ref{metrics} and \ref{power}, we chose $C = 20$ and $n = 300$, because for $\gamma = 0$ these parameters yield empirical power within $0.8-0.9$, which is a typical range used in cluster randomized trials.

\textbf{Network mixing.}\quad  In each cluster pair, one cluster is randomly assigned to treatment and the other is not.  The mixing parameter $\gamma$ can be expressed in terms of the entries in the adjacency matrix, $\mathbf{A}$, and the treatment assignment of clusters:

\begin{align}
\gamma :&= \frac{\sum_{ij}A_{ij}\left(1-\delta(r_i, r_j)\right)}{\sum_{ij}A_{ij}} \\
&= 1-\frac{1}{2m}\sum_{ij}A_{ij}\delta(r_i, r_j). \label{gamma_equation}
\end{align}

\begin{figure}[H]
\centering
\includegraphics[width=15cm]{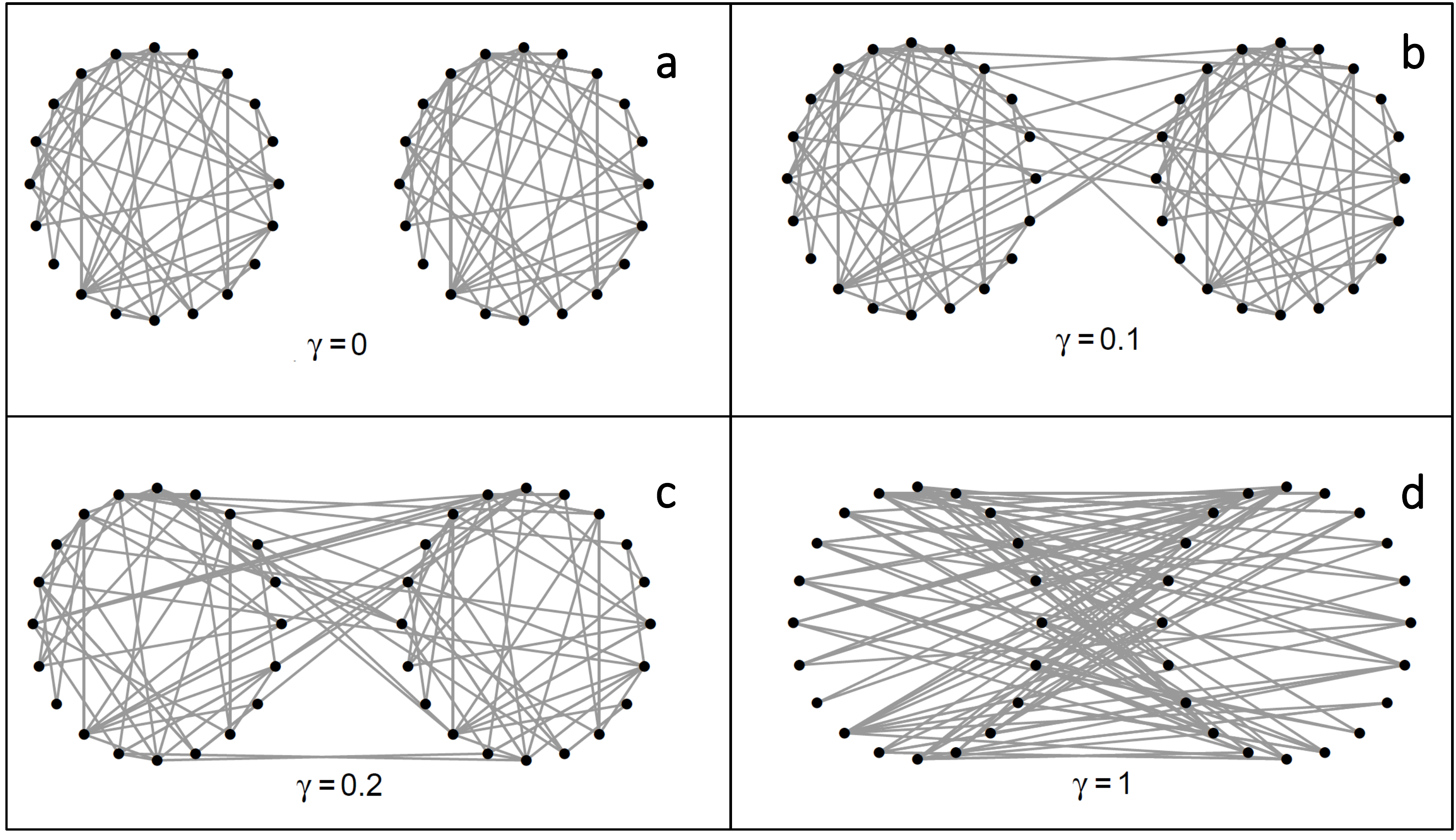}
\caption{A diagram showing two clusters with various proportions of mixing.}
\label{mixing_diagram}
\end{figure}

Here, $m$ is the total number of edges in the study, $r_{i}=1$ if node $i$ is in the treatment arm and $r_{i}=0$ otherwise, and $\delta(a,b)$ is equal to 1 when  $a=b$ and $0$ otherwise. This definition of between-cluster mixing is closely related to the concept of \textit{modularity}, used extensively in network community detection (see supplementary material \texttt{S2}).  If $\gamma=0$, the two clusters share no edges with each other. If $\gamma=1/2,$ there are as many edges reaching across two clusters as exist within them. Finally, if $\gamma=1,$ edges are only found between clusters, and the cluster pair network is said to be \emph{bipartite}. A schematic of network mixing is shown in Figure \ref{mixing_diagram}.

\begin{figure}[H]
\centering
\includegraphics[width=15cm]{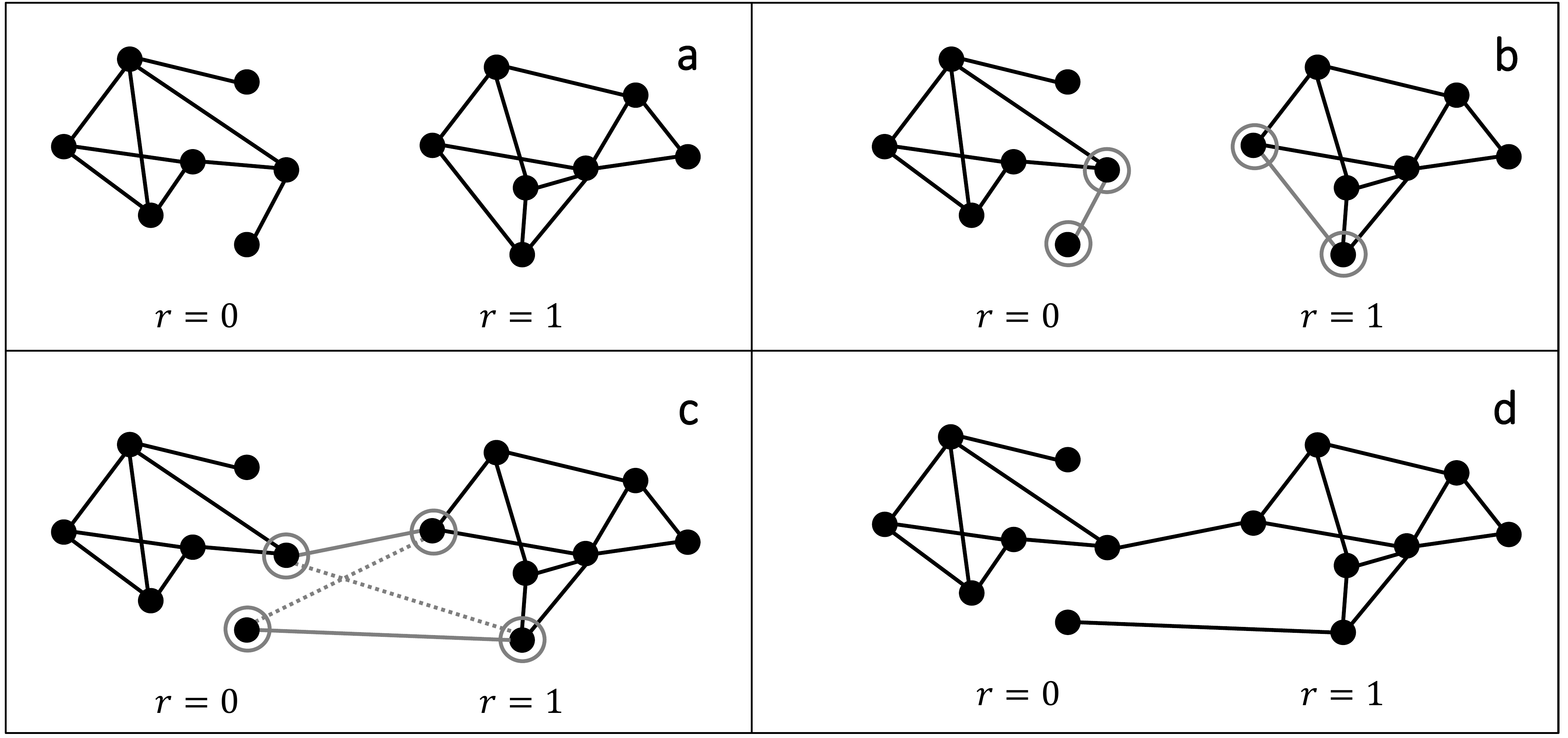}
\caption{Degree-preserving rewiring is performed by selecting an edge within each cluster, and swapping them to reach across the cluster pair.  The dashed gray lines represent another way the edges could have been rewired while still preserving degree; either rewiring is chosen with equal probability.}
\label{rewiring_diagram}
\end{figure}

\textbf{Network rewiring.}\quad  We first simulate two random networks from the same network model and with the same number of edges, each corresponding to a cluster in a pair of clusters. Then, we randomly select one edge from each cluster in the pair and remove these two edges. Finally we create two new edges among the four nodes such that the two edges reach across the cluster pair. This process is called \emph{degree-preserving rewiring}\cite{newman2003_1} because it preserves the degrees of all the nodes involved. The process is depicted in Figure $\ref{rewiring_diagram}$. We repeat the rewiring process until proportion $\gamma$ of the total edges are rewired. The result is a single cluster pair in our simulated CRT, and the pair-generating process is repeated until we have generated our target number of cluster pairs.

\textbf{Infectious spread.}\quad Compartmental models assume that each node in a population is in one of a few possible states, or compartments, and that individuals switch between these compartments according to some rules. Although more realistic models include more states\cite{vankleef2013}, we will assume for simplicity that nodes are in only one of two states: uninfected but susceptible ($S$), and infected and contagious ($I$).  We assume that the network structure of each cluster pair represents the possible transmission paths from infected nodes to susceptible ones.

Let $I_{irct}$ represent the infectious status for node $i$ in treatment arm $r=\{0,1\}$ and cluster pair $c=1,...,C$ at discrete time $t=1,...,T_c$, with $I_{irct}=1$ if the node is infected and $0$ otherwise.  We define $r=0$ if node $i$ is in the control arm, and $r=1$ if $i$ is in the treatment arm.  Let $I_{rct}:=\langle I_{irct}\rangle$ represent the proportion of infected nodes in cluster pair $c$ at discrete time $t$. At the beginning of the study, 1\% of individuals in each cluster is infected, i.e. $I_{rc0}=0.01$.  For each time step $t$, each node $i$ selects $q_{i}$ network neighbors at random, and infects each one with probability $p_{i}$.  Because different infectious diseases have different infectivity behavior, we study both unit and degree infectivity, or $q_i=1$ and $q_i=k_i,$ respectively.  We assume that the infection probability depends only on the treatment arm membership of each node $r_{i},$ thus $p_{i}=p_{r_{i}}$. Treatment reduces the probability $p_{r_i}$ of infection. If two clusters in a pair have the same infection rate, the treatment has no effect and $p_{r_{i}}=p$. This is the \emph{null hypothesis} under examination in our hypothetical study. When we simulate trials under the null hypothesis we set $p=0.30$ in every cluster.  The \emph{alternative hypothesis} holds if the treatment succeeds in reducing the infection rate, $p_{1}<p_{0}$. When we simulate under the alternative hypothesis, $p_{0}=0.30$ and $p_{1}=0.25$. The trial ends when the cumulative incidence of infection grows to 10\% of the population, i.e., when the cluster pair infection rate $\langle I_{ircT_c}\rangle=0.1$ for some time $T_c$.

\textbf{Analysis.}\quad At the end of the simulation, we test whether the treatment was effective by comparing the number of infections between treated and control clusters according to two analysis scenarios.  In real-world CRTs, the most efficient and robust way to compare the two groups depends on what information about the infection can feasibly be gathered from the trial. In some trials, surveying the infectious status of individuals is difficult, and therefore this information is only available for the beginning and end time points of the trial.  In others, the times to infection for each node are available.  In addition to what information is available, the researcher must choose a statistical test according to which assumptions they find suitable to their study. A \emph{model-based test} assumes that the data are generated according to a particular model, which can be more powerful than other tests if the model is true\cite{murray2004}.  Alternatively, a \emph{permutation test}\cite{good2001} does not make any assumptions about how the data were generated.  To show how to conduct an analysis suited to different scenarios based on available data, we analyzed our simulated trial using two different sets of assumptions. In Scenario 1, we assume that outcomes are only known at the end of the trial, and perform a model-based test.  In Scenario 2, we assume that the time to each infection is known, and perform a permutation test. We show that the results of the simulation are qualitatively similar under both scenarios.  (Note that it is possible to use a permutation test for Scenario 1 or a model-based test for Scenario 2, which would create two new analyses.)

\underline{Scenario 1:} The \emph{log risk ratio} is the logarithmic ratio of infected individuals in the treatment clusters to the control clusters at the end of study.  For simulation $m$, let $I^{(0)}_m:= \langle\log\frac{I_{0cT_c}}{I_{1cT_c}}\rangle=\langle\log I_{0cT_c}-\log I_{1cT_c}\rangle$ be the difference in the number of infections between two clusters in a pair averaged over each of the $C$ cluster pairs at the trial end $T_c$. The simulation was repeated 20,000 times under the null hypothesis and cutoff values $I^{*}_{2.5}$ and $I^{*}_{97.5}$ were established such that $P(I^{*}_{2.5}<I^{(0)}_m<I^{*}_{97.5})=\alpha$ for significance level $\alpha=.05$. We repeated this process under the alternative 20,000 times, and the proportion of these trials with statistics $I^{(A)}_m$ more extreme than $(I^{*}_{2.5}, I^{*}_{97.5})$ is the simulated power or empirical power.

\underline{Scenario 2:} We pool the individual infection times for the treatment arm and the control arm, and summarize the difference between the two arms' infection times using an appropriate statistic (e.g. the logrank statistic\cite{harrington2005}). The permutation test is performed by comparing the observed logrank statistic to the distribution of log-rank statistics when the treatment labels are permuted, or switched, for each cluster pair. The $p$-value for this analysis is the proportion of times the log-rank statistic with the real labels is more extreme than the permuted log-rank statistics. Because the permutation test is computationally expensive, this entire process is repeated 2,000 times, and we calculate the proportion of permutation p-values below 0.05, which is the empirical or simulated power.

We also compared this formulation to traditional methods, for which we take the formulas in Hayes and Bennett\cite{hayes1999} to be representative.  In this calculation, power is a function of number and size of clusters, the expected log risk ratio, and the expected average pairwise correlation of outcomes within each cluster (ICC).  The value from the ICC must be assumed beforehand or estimated in a small pilot study.  To compare this approach with our simulation design, we assumed that the ICC took on a range of plausible empirical values $0.003-0.06$ reported in the literature\cite{eldridge2012book}\cite{turner2005}.  For more details, see supplementary material \texttt{S4}.

{\large\textbf{Results}}

We begin by showing the effect of the mixing parameter $\gamma$ on the infection risk ratios between treated and untreated clusters. The means and standard deviations of simulated risk ratios observed under Scenario 1 are presented in Figure \ref{metrics}.

For both kinds of infectivity, neither the heavy-tailed degree distribution of the BA network nor the within-cluster community structure of the SBM network dramatically impacts the differences between the proportion of infections in the treated and controlled clusters in each pair (top row) compared to the ER network.  The differences between the risk of infections in the treated and untreated cluster pairs decreases as mixing increases, and reverses direction when $\gamma>1/2.$ This is expected because for this range of between-cluster mixing, infected individuals in the treatment cluster are more likely to contact members of the untreated cluster and vice versa, which is unlikely in practice but is included here for completeness. In almost all cases, the variation in the simulated studies' average log risk ratio decreases uniformly as $\gamma$ increases, which suggests that increasing the amount of mixing across communities results in less variation in the average rate of infections. However, the BA network is an exception. Under degree infectivity, when individuals can infect everyone to whom they are connected in a single time step, an infected node with large degree may spread its infection to each of its contacts at a single time point, which can cause a very fast outbreak. However, highly-connected individuals are rare, so in this case outbreaks are large but infrequent, increasing the variation in observed differences between treated and untreated clusters. This variation means that more clusters are required to estimate the average treatment effect with any precision. In other words, rare outbreaks make it harder to distinguish whether differences between the treatment arm and control arm are due to treatment or to a chance outbreak occurring in either arm. Therefore, under degree infectivity, the BA network results in less power than the SBM or ER networks, which shows that within-cluster network structure can impact the power to detect treatment effects in CRTs for certain kinds of infections.

For the two analysis scenarios described in Methods, we can directly estimate empirical power as the proportion of simulations resulting in the rejection of the null hypothesis at the $\alpha=0.05$ level under the alternative for a range of mixing values $\gamma$. Our results, as well as a comparison with the standard approach, are summarized in Figure \ref{power}.

In all settings, power is lowest when $\gamma\approx1/2,$ with approximately the same number of edges between clusters as within them. Scenarios 1 and 2 (the top and bottom rows, respectively) show few differences from one another, which suggests that the two strategies for significance testing tend to give qualitatively similar results. Unit infectivity (lefthand column) shows no differences in power among network types. This is not the case for degree infectivity (righthand column), in which the BA network shows less power than the other networks, for the reasons discussed above.  Finally, the gray bars indicate that when no mixing is present, standard power calculations are conservative for all network types we studied, and no sample size adjustment may be needed.  However, moderate to severe between-cluster mixing can greatly overestimate expected power.  In the case of the BA network and degree infectivity, the standard approach always overestimates trial power.

\textbf{Size and number of study clusters.} Our results so far have shown how power in CRTs is affected by between-cluster mixing,  within-cluster structure, and infectivity. Next, we show how power relates to other trial features, namely the size and number of clusters, $n$ and $C$, respectively. The results are qualitatively similar for Scenarios 1 and 2, and the results shown in Table \ref{sensitivity_analysis} are for Scenario 1. The table shows results for each combination of a range of cluster sizes $n=\{100,300,1000\}$ and numbers $C=\{5,10,20\}$ as a $3\times3$ grid of pairs of cells. Each cell pair is a side-by-side comparison of results for unit infectivity (lefthand cell) and degree infectivity (righthand cell). Each cell shows simulated results for within-cluster structure (columns) as well as amount of between-cluster mixing (rows). Considering the case of $C=10, n=300$ (the middle-most cell pair), we notice a few trends. We see that increasing mixing (looking down each column) decreases power in all cases. We can directly compare the two types of infectivity (comparing cells in the pair), and see that all the entries are similar except for the BA network (middle column). For BA networks, power is much lower for degree infectivity spreading compared to unit infectivity. This suggests that CRTs with network structure similar to BA networks can have substantially less power when the infection spreads in proportion to how connected each node is. Finally, we may compare studies of differing cluster numbers and sizes (comparing cell pairs), and see qualitatively similar results: in each case, more or larger clusters in the study (cell pairs further down or right) result in more power overall.  When power is very high (bottom-right cell pair), within-cluster structure affects results less.  Therefore, careful consideration of expected power is most important when trial resources are limited, which is often the case in practice.

\begin{table}[H]
\centering
\includegraphics[width=17.5cm]{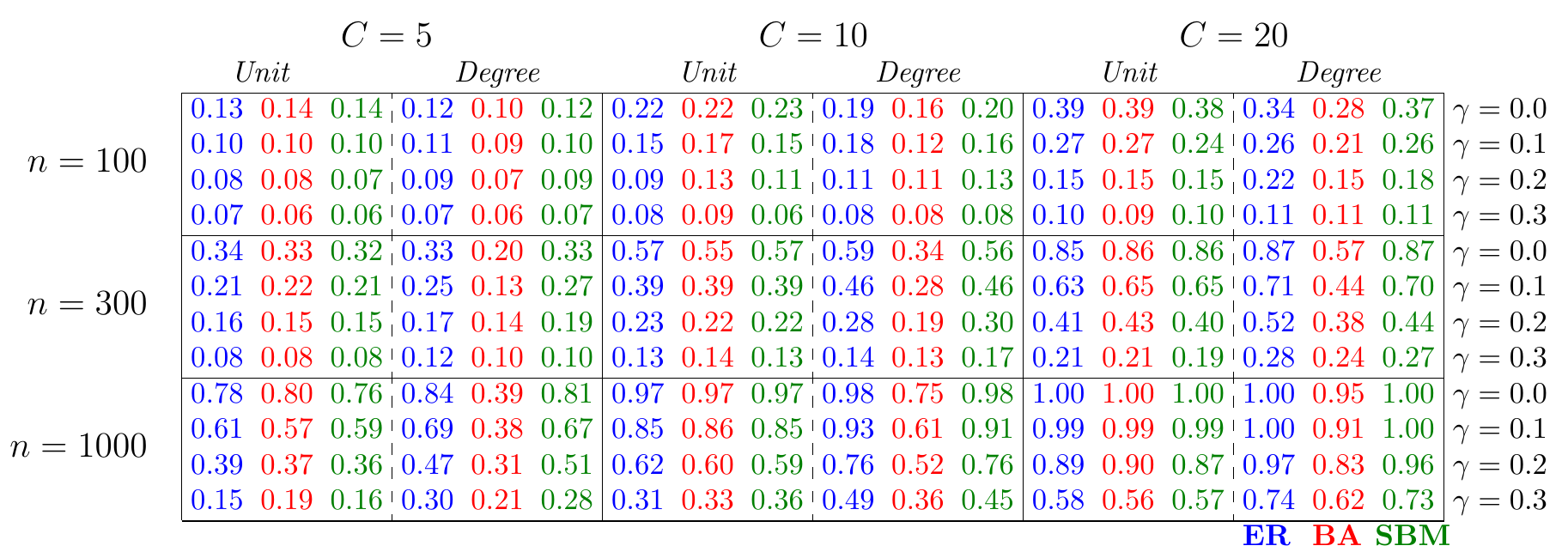}
\caption{Experimental power in our simulation framework for different sizes and numbers of cluster pairs, $n$ and $C$, respectively, for Scenario 1. Each cell shows output for $3,000$ simulations of each combination of $n$ and $C$, all three within-cluster structures, various values of  mixing parameter $\gamma$, and both unit and degree infectivity. The results are similar for Scenario 2.}
\label{sensitivity_analysis}
\end{table}

\textbf{Real-world data and the extent of mixing.}  Finally, we show how our mixing parameter can be estimated using data in the planning stages of a hypothetical CRT.  Sometimes the entire network structure between individuals in a prospective trial is known beforehand, such as the sexual contact network on Likoma Island\cite{helleringer2007}. In this case, between-cluster mixing can be estimated using Equation \ref{gamma_equation}.  In other trials, perhaps only partial information is known, like the degree distribution\cite{barabasi1999} and/or the proportion of ties between clusters.  In this case, clusters can be generated that preserve partial network information such as degree distribution\cite{goyal2014}\cite{molloy1995}, and degree-preserving rewiring can be performed until proportion $\gamma$ of ties between clusters is observed, where this quantity is estimated from the network data, if possible.

The structure of calls between cell phones is often persistent over time\cite{hidalgo2008} and indicative of actual social relationships\cite{eagle2009}.  We use a network of cell phone calls as a proxy for a contact network, we use our definition of between-cluster mixing to estimate the amount of mixing between hypothetical clusters. The dataset consists of all the calls made between cellphones of a large mobile carrier within a quarter year.  Individual phone numbers were anonymized, and we only report results for the number of individuals and calls within or between zip codes.

The dataset contains phone calls originating from $Z=3806$ different zip codes, and we define a cluster as a collection of zip codes that are spatially close to one another.  Because zip codes are numerically assigned according to spatial location, we assume that zip codes that are numerically contiguous to each other are also close to each other spatially. Therefore, zip code $z=1,...,Z$ assigned to cluster $c_z=1,...,2C$ is
\begin{align}
c_z:= \left\lceil \frac{z}{Z}2C \right\rceil
\end{align}
where $2C$ is the total number of clusters in the trial, and $\left\lceil \cdot \right\rceil$ is the ceiling function.  Once the number of clusters $2C$ is specified, clusters may be paired, with one cluster in each pair randomized to a hypothetical treatment, and the other to the control condition.

\begin{figure}[H]
\centering
\includegraphics[width=14.5cm]{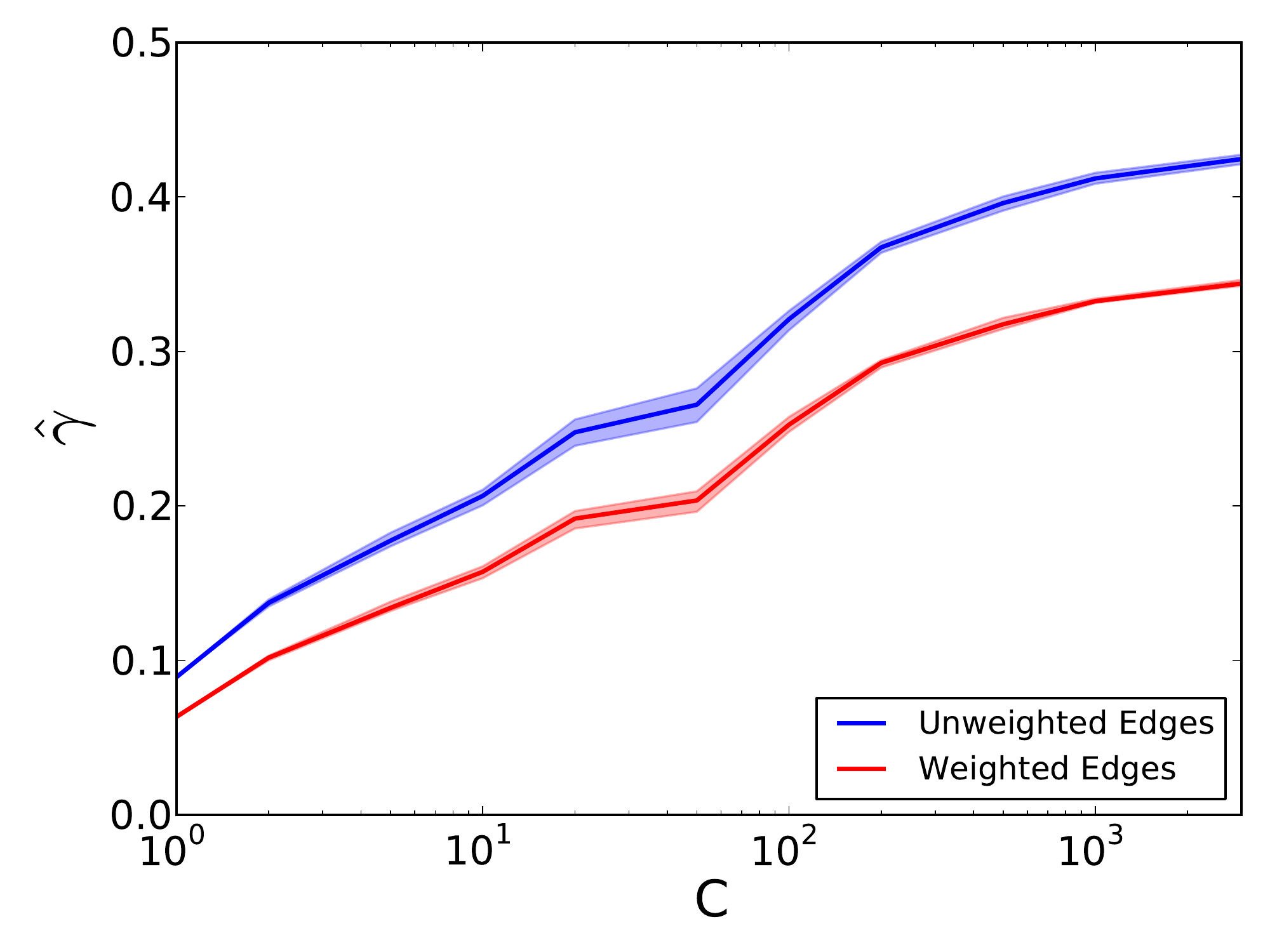}
\caption{A log-linear plot displaying empirical values of mixing parameter $\gamma$.  The $y$ axis shows the mean and $(2.5, 97.5)$ quantiles of these estimates.  The $x$ axis in each panel corresponds to a range of cluster numbers $C$.}
\label{application_function}
\end{figure}

Next, we estimate mixing parameter $\gamma$ for this dataset. We consider two definitions for the number of edges shared between individuals, one in which they are unweighted and one in which they are weighted by the number of calls between them.  We consider two definitions for an edge $A_{ij}$ between individuals $i$ and $j$, belonging to clusters $c_i$ and $c_j$ respectively.  The number of calls between $i$ and $j$ over the period of investigation is defined as $d_{ij}.$  For Definition 1, we assume and edge exists between the two individuals if they have called each other at least once, $A_{ij}=\mathbb{I}(d_{ij}\geq 1)$, and otherwise no edge exists between them $A_{ij}=0$.  For Definition 2, we assume an edge between them may be weighted by the number of total calls made between them, $A_{ij}=d_{ij}$.  Using both definitions, we found the degree distribution of each cell phone to be heavy-tailed (see supplementary material \texttt{S5}). For a range of numbers of cluster pairs $C$, we cluster all $Z$ zip codes into $2C$ clusters, and randomize one cluster in each pair to a hypothetical treatment, and the other to a control. For $200$ randomizations, we calculate the between-cluster mixing parameter $\gamma$ using Equation \label{gamma_equation}.  We examine the relationship between $\gamma$ and the number of clusters $C$.  The mean and $(2.5, 97.5)$ percentiles of these estimates as a function of the number of clusters number $C$ are shown in Figure \ref{application_function}.

Figure \ref{application_function} displays a number of distinct trends.  As the number of clusters increases, fewer of the total zip codes are included in each cluster, and the number of calls between clusters increases.  This means that individuals are more likely to call others in zip codes geographically closer to them, which has been confirmed in other phone communication networks\cite{onnela2011}.  Between-cluster mixing unweighted by the number of calls (blue) results in higher estimates of $\gamma$ than weighted (red), which means that when individuals call others outside their cluster, they tend to call those people less than others they call within their cluster.  There is significant between-cluster mixing for all values of $C$, implying that between-cluster mixing would significantly decrease the power of a trial that assumes each cluster to be independent ($\gamma=0$).  Furthermore, as the number of clusters increases, the average cluster size decreases, and mixing reaches a maximum of $\gamma=0.45$.  Extrapolating from our simulation framework, power could be reduced dramatically in this case.

{\large\textbf{Discussion}}

Before conducting a trial, it is important to have an estimate of statistical power in order to assess the risks of failing to find true effects and of spurious results. If individuals belong to interrelated clusters, randomly assigning them to treatment or control may not be a palatable option, and CRTs can be used to test for treatment effects. Power in CRTs is known to depend on the number and size of clusters, as well as the amount of correlation within each cluster. However, within-cluster correlation structure is often measured by a single number and clusters are usually assumed to be independent of one another.  Unfortunately, these assumptions can produce misleading estimates of power.

To investigate this problem, we studied the effects of complex within-cluster structure, a measure of between-cluster mixing strength, and infectivity on power by simulating a matched-pairs CRT for an infectious process.  We simulated a collection of cluster pairs as a network, controlling the proportion of edges shared across each pair.  We then simulated an $SI$ infectious process on each cluster pair, with one cluster assigned to treatment and the other assigned to control. The effect of treatment in this simulation lowered the probability that an infected individual succeeds at infecting a susceptible neighbor. We also considered two types of infectivity: unit and degree.

We found that between-cluster mixing had a profound effect on statistical power, no matter what network or infectious process was simulated.  As the number of edges shared across clusters in different treatment groups increased to $1/2$, on average the two clusters were nearly indistinguishable, and thus power fell to nearly zero.  This is not surprising, but most power calculations assume clusters are independent, and this issue is usually left unaddressed.  We compared these findings to the ICC approach, and found it will significantly underestimate expected power if the extent of between-cluster mixing is moderate to severe.

The effect of within-cluster structure was more nuanced.  For degree infectivity, the spread of infection was less predictable if the network contained some highly-connected nodes, due to the variation in and strong effects of these hubs becoming infected.  We did not observe this level of variability for networks without highly-connected hub nodes. We also did not observe this level of variability for unit infectivity, regardless of how many hubs were present in the network.  Taken together, we found that for the network structures we studied, within-cluster structure had a significant impact on power only when the infectious process exhibited degree infectivity.  The effect of within-cluster structure and between-cluster mixing on statistical power are qualitatively similar for a range of cluster sizes and numbers, although (as is well known) an increase in either results in more power overall.

Our simulation framework can be used to estimate power before an actual trial.  If partial or full network information is available, it can be used to simulate an infectious processes using a compartmental model, and analyze the resulting outcomes as we have described.  We demonstrated how to estimate between-cluster mixing using a dataset composed of cellphone calls from a large mobile carrier, which are taken to represent a contact network.  For a hypothetical prospective trial on the individuals in this dataset, we defined a cluster as a group of individuals within a collection of contiguous zip codes.  We then grouped clusters into pairs, randomly assigned one cluster in each pair to a hypothetical treatment condition and the other to a control, and estimated mixing parameter $\gamma$ for each simulation. We found substantial between-cluster mixing for all choices of cluster numbers, and mixing increased when clusters were chosen to be more numerous but smaller. Estimates of between-cluster mixing ranged from moderate to severe, regardless of whether the estimation adjusted for the frequency of calls or not.

Our study invites several investigations and extensions.  First, we have employed restrictively simple network models and infectious spreading process, and more nuanced generalizations are available.  While our work shows how infectious spreading and complex structure can affect expected results in CRTs, more specific circumstances require extensions with more tailored network designs and infection types for power to be properly estimated.  Second, we have focused our attention on matched-pair CRTs, and our framework should be extended to other CRT designs used in practice\cite{eldridge2012book}. Third, these findings should be replicated in data for which both network structure and infectious spread are available.

\renewcommand{\figurename}{Supplementary Figure}
\renewcommand{\tablename}{Supplementary Table}
\setcounter{figure}{0}
\setcounter{table}{0}

\section*{Acknowledgements}
\vspace{-.5cm}
We thank Telenor Research and Senior Data Scientist Kenth Eng\o-Monsen for supplying the telecom dataset, and helping with initial processing and answering questions related to the dataset.  We also thank the National Institutes of Health for funding support for PCS (NIH-5T32ES007142-32, PI Coull).

\newpage
\begin{center}
{\huge Supplementary Material}
\end{center}
\vspace{1cm}

In this supplement, we provide additional details for a few topics discussed in the main paper.  Section \texttt{S1} demonstrates a simple approach to modeling infectious spread with between-cluster mixing using ordinary differential equations, and compares this result to the simulation approach introduced in the paper.  Section \texttt{S2} describes the stochastic blockmodel and provides details for the specific model we used in our paper.  Section \texttt{S3} connects our definition of between-mixing parameter $\gamma$ with a common metric used in applications of network science.  Section \texttt{S4} describes how the Intracluster Correlation Coefficient is defined, and we show estimates of this quantity for our simulations.  Finally, Section \texttt{S5} shows the degree distribution for the empirical cell phone network, with discussion.

\section*{\texttt{S1}:  Ordinary Differential Equation approach to epidemic spreading with between-cluster mixing.}

One of the most common approaches to investigating the spread of an epidemic on networks is Ordinary Differential Equations (ODEs)\cite{anderson1992book}\cite{pastorsatorras2014}.  ODEs are functions of a variable in terms of its derivatives.  Compartmental models for epidemic spread can use ODEs to specify the rate of change for individuals in terms of others.  A common assumption used to specify ODEs for epidemic spread is \emph{mass action}, in which the spread of an infection depends only on the proportion of individuals in each compartment. For example, an $SI$ compartmental model assumes that individual $i$ is either infected ($I_i(t)=1$) or not infected but susceptible ($S_i(t)=1$) at any time $t$.  These two statuses are mutually exclusive, and $S_i(t) = 1-I_i(t)$.  An ordinary differential equation that assumes mass action would specify the change in the total proportion of infected individuals $I(t) := \langle I_i(t) \rangle$ in terms of the infected proportion $I(t)$ at time $t$.  If we assume mass action, we may model the rate of infectious growth in an $SI$ compartmental model as proportional to the proportion of infected individuals multiplied by the proportion of susceptible individuals:

\begin{align}
\frac{dI(t)}{dt} = p S(t)I(t) = p (1-I(t))I(t) \label{ODE_eq}
\end{align}

In this paper, we consider a collection of $c=1,...,C$ cluster pairs, with one cluster in each pair assigned to the treatment condition $r=1$ and the other to control $r=0$.  Furthermore, we assume that clusters are mixed according to mixing parameter $\gamma$,  For the $SI$ compartmental model, $I_{irc}(t)=1$ if individual $i$ is infected and $0$ otherwise.  We may assume that the spread of an infection across the network pair is a mass action ODE as above, with a simple modification.  Let $I_{rc}(t)=\langle I_{irc}(t)\rangle$ represent the proportion of infected nodes in cluster pair $c$ at discrete time $t$.  Individual $i$ may contact an individual $j$ in the opposing cluster with probability $\gamma$.  In this case, the probability of a successful infection requires that $i$ is suspectible and $j$ is infectious.  Mass action dictates that the rate of change for each cluster depends only on the proportion of individuals in each infectious status for either cluster, which is now sum of ODEs weighted by mixing parameter $\gamma$:
\begin{align}
\frac{\partial I_{0c}(t)}{\partial t} &= \left[(1-\gamma) I_{0c}(t) p_0 + \gamma I_{1c}(t) p_1 \right](1-I_{0c}(t)) \label{diff_eq1}\\
\frac{\partial I_{1c}(t)}{\partial t} &= \left[(1-\gamma) I_{1c}(t) p_1 + \gamma I_{0c}(t) p_0 \right](1-I_{1c}(t)) \label{diff_eq2}
\end{align}

According to Supplementary Equations \ref{diff_eq1} and \ref{diff_eq2}, if $\gamma=0$, the rate of infection in each cluster is identical to Supplementary Equation \ref{ODE_eq}.  As $\gamma$ approaches $1/2$, the difference in the proportion of infected individuals in the two treatment arms decreases to no difference.

\begin{figure}[H]
\centering
\begin{multicols}{2}
\includegraphics[page=1,width=8cm]{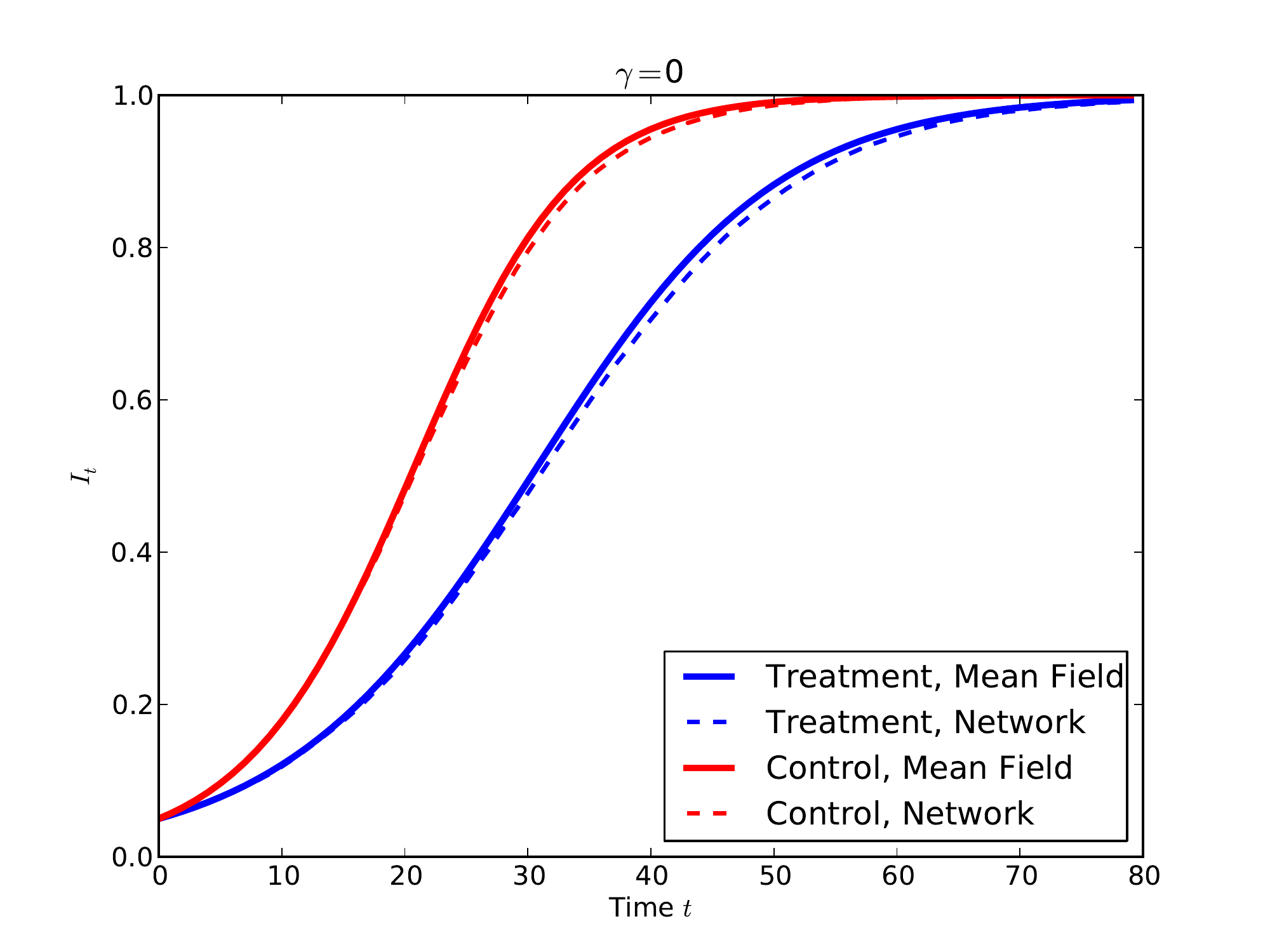} \\
\includegraphics[page=3,width=8cm]{supplementary_figure_1.pdf} \\
\includegraphics[page=2,width=8cm]{supplementary_figure_1.pdf} \\
\includegraphics[page=4,width=8cm]{supplementary_figure_1.pdf}
\end{multicols}
\caption{The proportion of infections over time.  The solid line is the mass action rate equation, and the dashed lines are the mean of simulations of an infectious process on a complete (fully-connected) network.  The infectious process was simulated for $\gamma=\{0.0,0.1,0.2, 1\}$, matching Figure 5.  As $\gamma$ approaches $1/2$, the difference in infection rates in two clusters in a pair decreases, demonstrated by the red and blue curves approaching each other.  When $\gamma=1,$ the relative rates of infections switch.}
\label{over_time}
\end{figure}

The ODE approach is quite comparable to the stochastic approach we chose for the paper.  To show this, we created network clusters with every node connected to each other in the cluster, performed degree-corrected rewiring, simulated an infectious processes with unit infectivity on the pair according to the paper, and averaged the proportion of infections at each time step. Supplementary Figure \ref{over_time} shows the infection rates over time for a range of mixing values $\gamma=\{0.0,0.1,0.2, 1\}.$ The solid lines shows the average of the network simulations. The dashed lines show the a numerical solution to Supplementary Equations \ref{diff_eq1} and \ref{diff_eq2}. The two are comparable, suggesting that differential equations and network simulations can approximately interchangeably describe the same infectious process.

Where the differential equation approach assumes individuals contact everyone in the population, infections spreading through fixed networks only allow contact through existing edges. This \emph{redundant contact effect}\cite{zhou2006} causes infections through networks to be slightly slower, also  observable in Supplementary Figure \ref{over_time}.

\section*{\texttt{S2}:  Modularity and Between-Mixing Parameter $\gamma$}

Our definition of between-mixing parameter $\gamma$ (Equation 2) has a convenient interpretation in terms of findings in network science.  Modularity $Q$ is a measure of how well the individuals in a network and their relationships fit into mutually exclusive groups\cite{newman2011}.  For CRTs, we assume the natural groupings to be the two treatment arms.  If $Q=1$, all edges exist within treatment arms.  If $Q=-1,$ all edges are between the two treatment arms.  The definition of modularity is written in the same terms as $\gamma$:

\begin{align}
Q:=\frac{1}{2m}\sum_{ij}\left(A_{ij}-\frac{k_ik_j}{2m}\right)\delta(r_i, r_j)
\end{align}

If the individuals between the two treatment arms have equal numbers of edges, $\sum_{ij}\frac{k_ik_j}{(2m)^2}\delta(r_i, r_j)=1/2,$ and $\gamma = 1/2 - Q.$  Therefore, if modularity can be computed, so can the mixing between the two treatment arms.  More generally, $\gamma$ is entirely a function of cluster structure matrix $\mathbf{A}$ and treatment assignments, so if an experimenter knows the structure of relationships among individuals in the study, they may calculate the estimate the amount of mixing between the two treatment arms.

\section*{\texttt{S3}:  Details on the Stochastic Blockmodel}

A stochastic blockmodel (SBM) is a probablistic network model, which means that the probability of an edge existing between nodes $i$ and $j$ is specified by probability $p_{i,j}$.  SBM assumes that each network node is a member of a exactly one block in a partition of $b$ blocks $\mathcal{B}=B_1,...,B_b$, and the probability $p_{i,j}$ of a connection between nodes $i$ and $j$ depends only on each node's block membership.  Denote the block membership of node $i$ as $B_i$. A probability matrix $P_{b\times b}$ describes all edge probabilities for a network, with $p_{i,j}=P_{B_i, B_j}$.

In our study, we imitated within-cluster community structure using a SBM.  We assume each cluster is comprised of blocks arranged in a triangular lattice structure.  Blocks of nodes may be thought of near each other in geographic location, and while most edges are contained within each block, blocks share a few edges according to a triangular spatial pattern.  We organized clusters into 10 equally-sized blocks, and individuals within each block are connected to others within their block such that average within-block degree is $\frac{9}{10}\langle k\rangle$.  For between-block connections, we also assume that each edge between members of blocks share a total between-block degree of $\frac{1}{10}\langle k\rangle$ with adjacent blocks according to the lattice structure, and no edges with all other blocks.  A diagram of this network ensemble is shown in Supplementary Figure \ref{spatial_SBM}.

\begin{figure}[H]
\centering
\includegraphics[width=11cm]{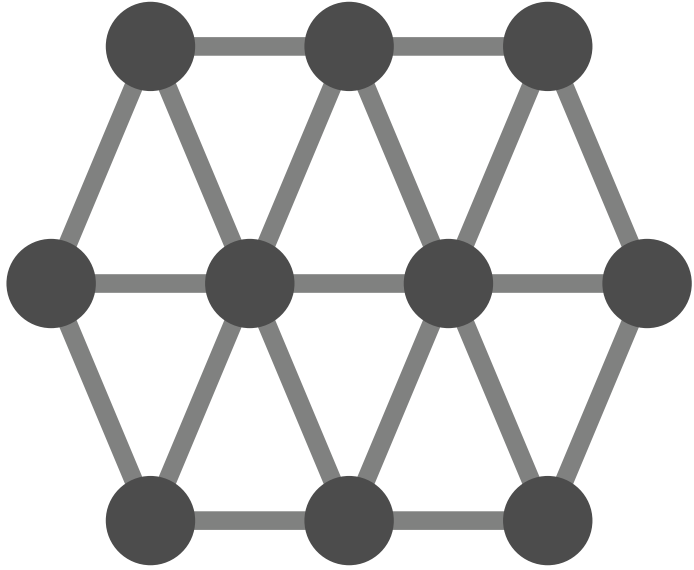}
\caption{10 communities or blocks within clusters were created according to the stochastic blockmodel, with a small probability of community ties in a triangular lattice.  Edge probabilities were selected to preserve the average degree of a random network.}
\label{spatial_SBM}
\end{figure}

\section*{\texttt{S4}:  The ICC}

The Intracluster Correlation Coefficient (ICC) is a measure of the average correlation between individual outcomes within a cluster.  The ICC assumes that the correlation is identical for all pairs of individuals within a cluster, and is constant across clusters.  The ICC can also be expressed as the ratio of between-cluster variance to the total outcome variance in the study\cite{kerry1998}.  In the case of binary outcomes, this value may be expressed as\cite{eldridge2012book}

\begin{align}
\text{ICC} &= \frac{\langle\pi_c(1-\pi_c)\rangle}{\langle\pi_c\rangle(1-\langle\pi_c\rangle)}
\end{align}

where $\pi_c$ is the proportion of infections in cluster $c$ and $\langle\cdot\rangle$ is the average over all clusters in a trial.  We calculated the ICC this value for each network ensemble and value of $\gamma$ in our simulations.  These results are shown in Supplementary Figure \ref{ICC_diagram}.

\begin{figure}[H]
\centering
\includegraphics[page=1,width=8.5cm]{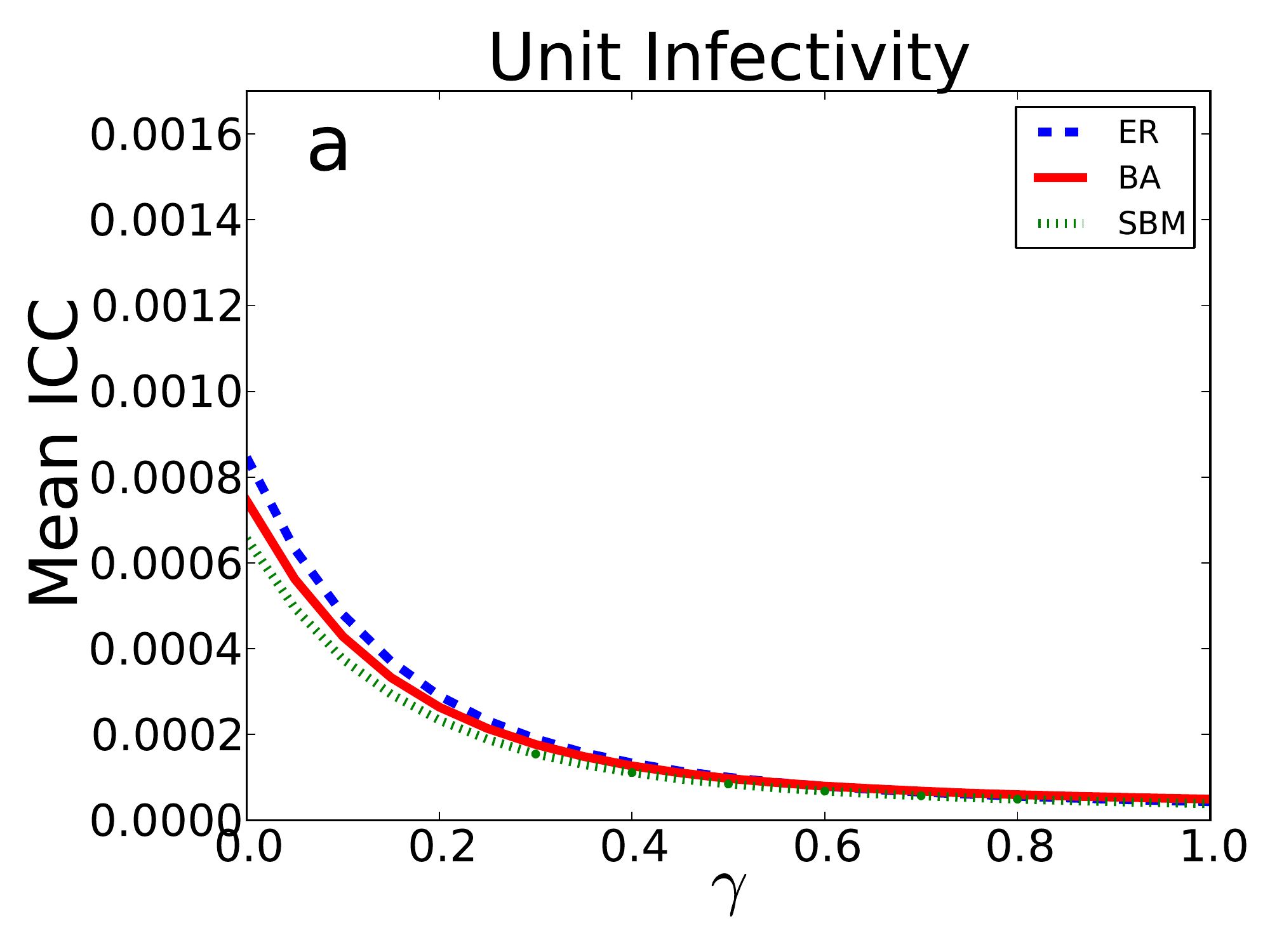}
\includegraphics[page=1,width=8.5cm]{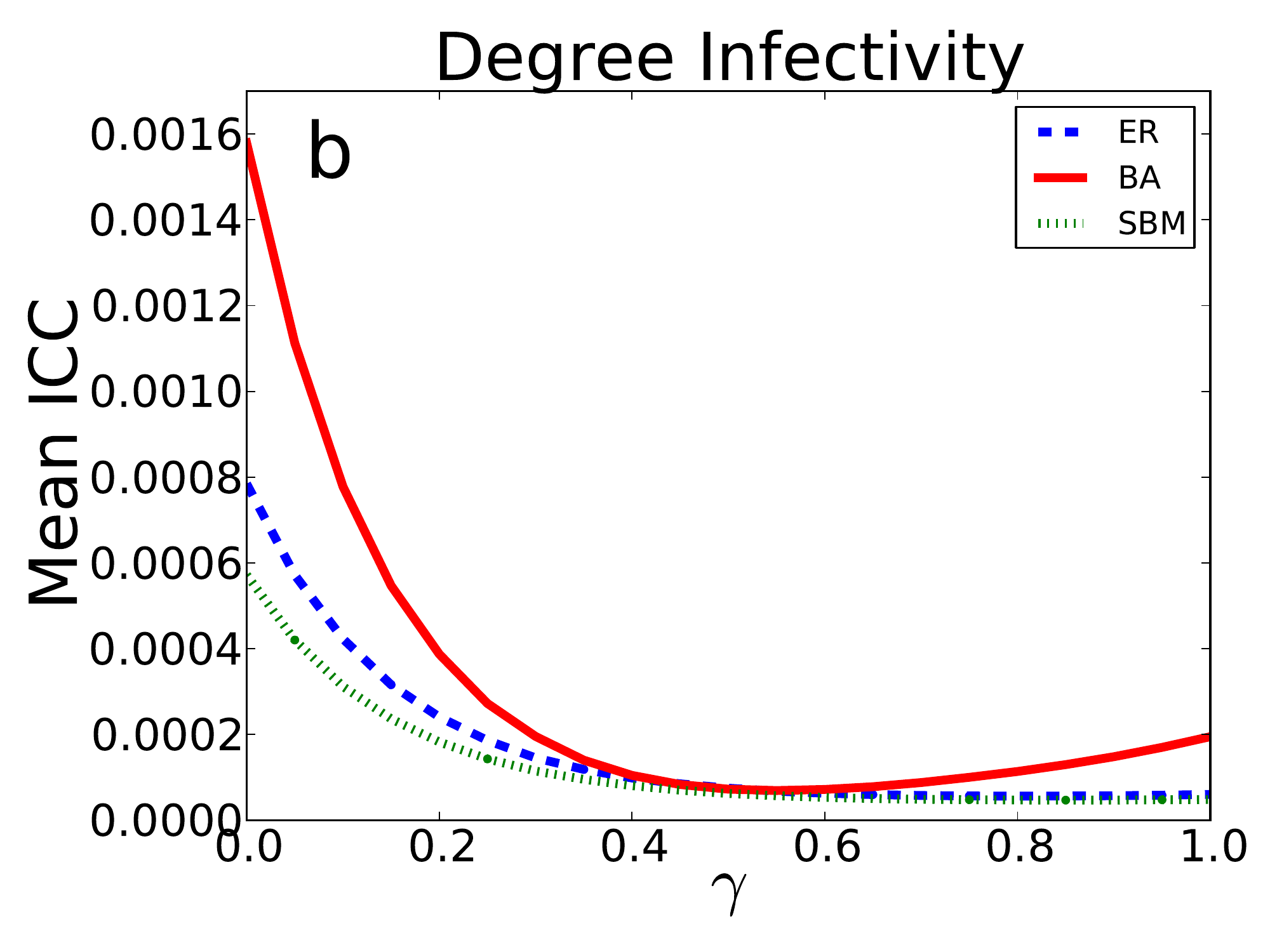}
\caption{ICCs from Scenario 1, averaged over all simulations.  ICC values are shown for unit infectivity (Panel \texttt{\textbf{a}}) and degree infectivity (Panel \texttt{\textbf{b}}), as well as each within-cluster structure and extent of between-cluster mixing specified in our simulations.}
\label{ICC_diagram}
\end{figure}

These values are quite low, but not very far from typical values\cite{turner2005} and lower values have been reported in actual trials\cite{eldridge2012book}.  These values for the ICC are low because in our design, the data is collected for each cluster pair when the average proportion of infections within each pair is 10\%, which results in relatively low variation in infection proportions for each cluster.

Like power, the relative value of the ICC depends on within-cluster structure, the amount of between-cluster mixing, and infectivity.  In the case of unit infectivity, the ICC shrinks as between-cluster mixing increases for all within-cluster structures.  However, in many power calculation formulas\cite{hayes1999}, lower values of ICC indicate increased power, not less.  This shows that even if sample size calculations account for within-cluster correlations as measured by the ICC, power can be reduced by other trial features, such as the extent of between-cluster mixing.

\section*{\texttt{S5}:  Degree Distribution for an Empirical Cell Phone Network}

The main paper specifies two definitions for an edge between callers in the cell phone network, which are, respectively, unweighted or weighted by the number of total number of calls made between each pair of callers.  The empirical degree distribution for both definitions are found in Supplementary Figure \ref{degree_distributions}.

\begin{figure}[H]
\centering
\includegraphics[page=1,width=8.5cm]{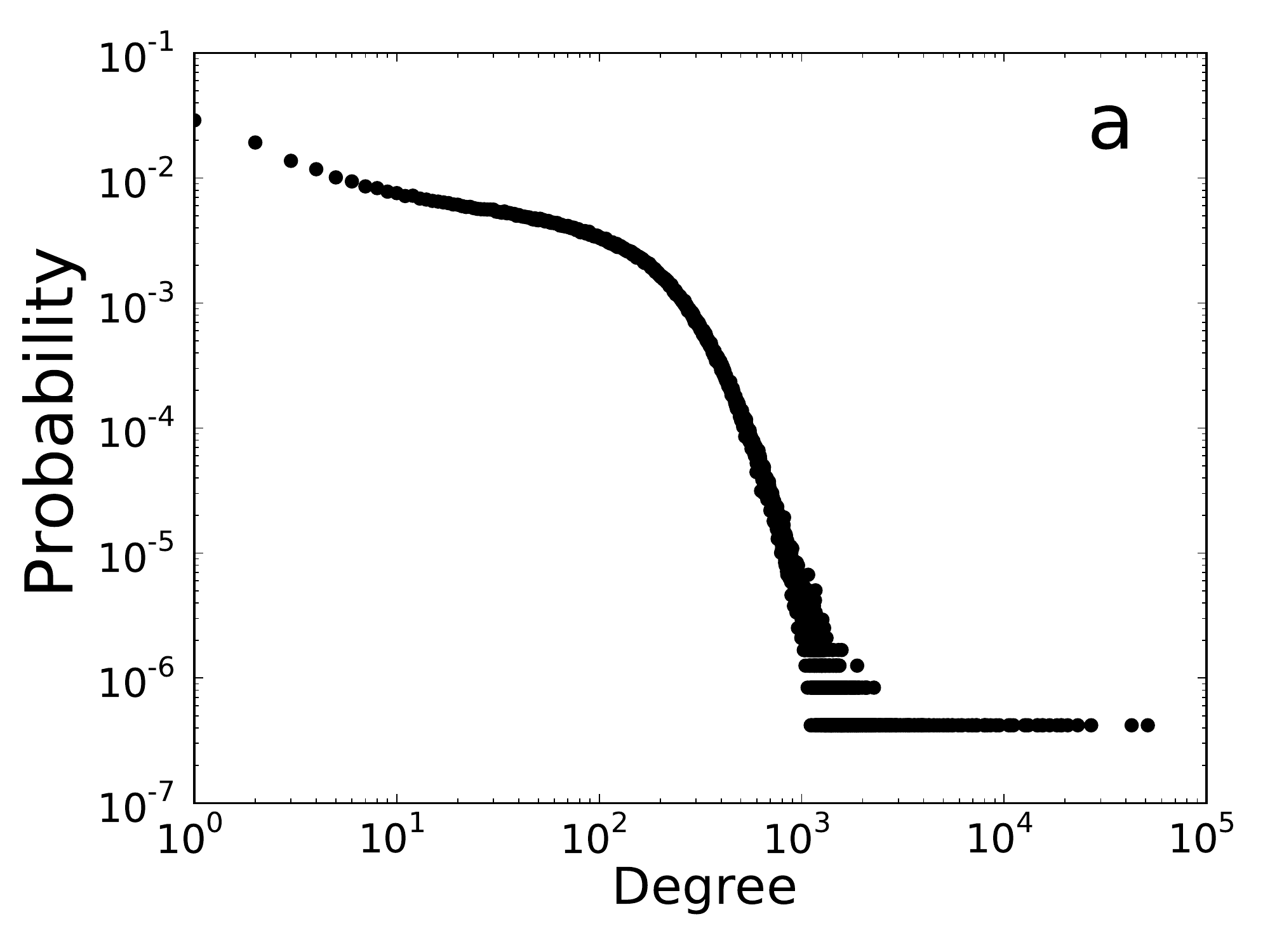}
\includegraphics[page=1,width=8.5cm]{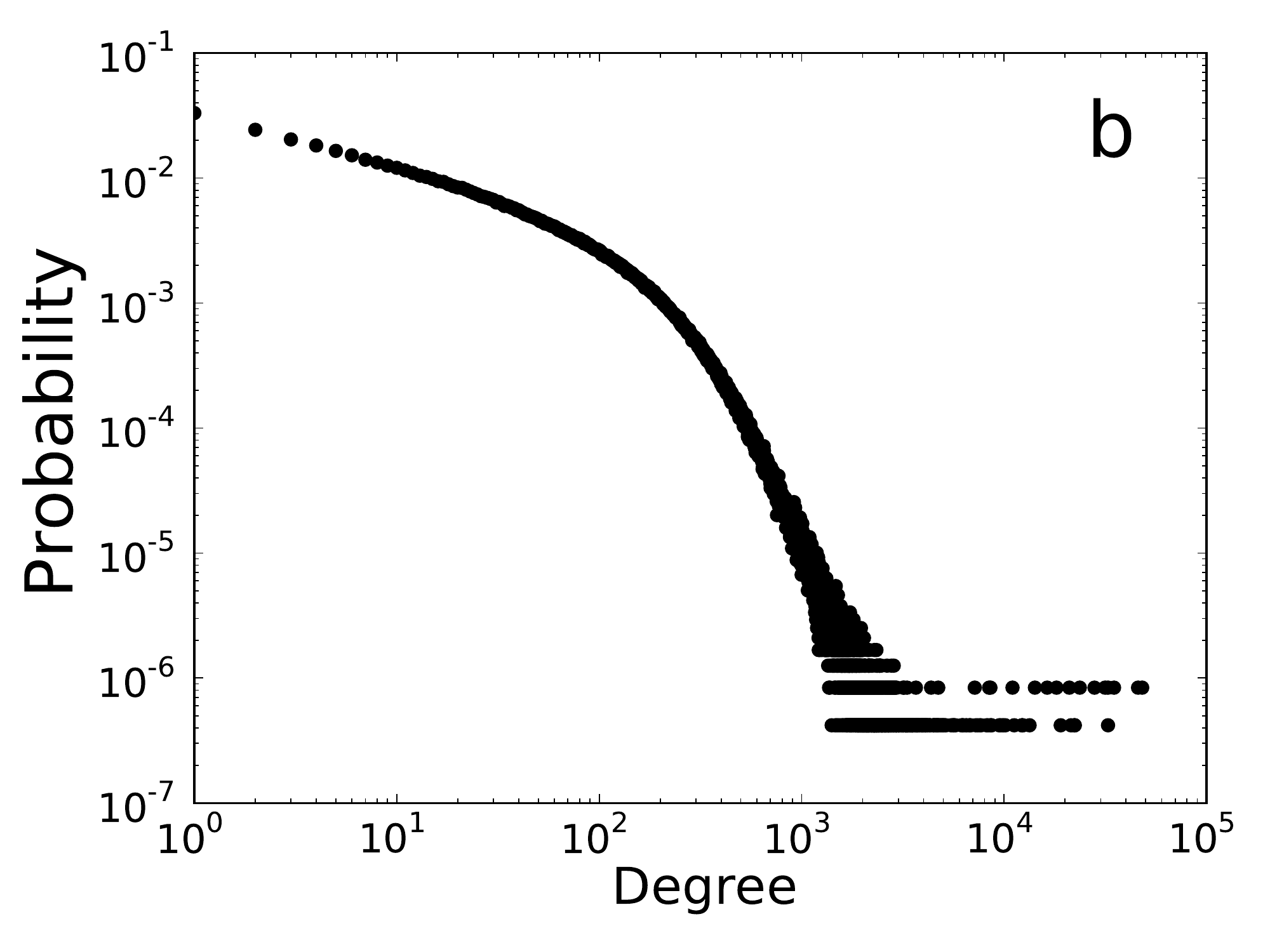}
\caption{The empirical degree distribution for the calling network dataset. Panel \texttt{\textbf{a}} corresponds to Definition 1 (unweighted), and Panel \texttt{\textbf{b}} corresponds to Definition 2 (weighted).}
\label{degree_distributions}
\end{figure}

Focusing on Panel \texttt{\textbf{a}}, we notice three distinct regimes.  The vast majority of callers make calls with $1-100$ others.  The distribution of those who call a large number $(100-1000)$ of others follows a nearly straight line on these log-log plots, which is indicative of a power-law for this segment.  Finally, a few singular callers are found to call a very large number $(>1000)$ of callers within the quarter. The general shape is similar for both the unweighted and weighted definitions. This degree distribution is in accordance to similar datasets analyzed in the literature\cite{onnela2011}.

\end{document}